# Modelling Temperature Variation of Mushroom Growing Hall Using Artificial Neural Networks

Sina Ardabili[1], Amir Mosavi[2,3], Asghar Mahmoudi[4], Tarahom Mesri Gundoshmian[5], Saeed Nosratabadi[6], and Annamária R. Várkonyi-Kóczy[2,7]

[1] Institute of Advanced Studies Koszeg, Koszeg, Hungary
[2] Institute of Automation, Kalman Kando Faculty of Electrical Engineering, Obuda University, Budapest, Hungary
amir.mosavi@kvk.uni-obuda.hu
[3] School of the Built Environment, Oxford Brookes University, Oxford OX30BP, UK
[4] Department of Biosystem Engineering, University of Tabriz, Tabriz, Iran
[5] Department of Biosystem Engineering, University of Mohaghegh Ardabili, Ardabil, Iran
[6] Institute of Business Studies, Szent Istvan University, Godollo 2100, Hungary
[7] Department of Mathematics and Informatics, J. Selye University, Komarno, Slovakia

**Abstract.**

The recent developments of computer and electronic systems have made the use of intelligent systems for the automation of agricultural industries. In this study, the temperature variation of the mushroom growing room was modeled by multi-layered perceptron and radial basis function networks based on independent parameters including ambient temperature, water temperature, fresh air and circulation air dampers, and water tap. According to the obtained results from the networks, the best network for MLP was in the second repetition with 12 neurons in the hidden layer and in 20 neurons in the hidden layer for radial basis function network. The obtained results from comparative parameters for two networks showed the highest correlation coefficient (0.966), the lowest root mean square error (RMSE) (0.787) and the lowest mean absolute error (MAE) (0.02746) for radial basis function. Therefore, the neural network with radial basis function was selected as a predictor of the behavior of the system for the temperature of mushroom growing halls controlling system.

Keywords: Agricultural production, Environmental parameters, Mushroom growth prediction, Machine learning, Artificial neural networks (ANN), Food production  Food security

## 1 Introduction

Nowadays, Due to issues such as population growth and limited agricultural resources including land and freshwater, the necessity of attention to new methods and efficiency in agricultural production, is quite evident [1]. A number of clinical studies in Japan and the United States of America have shown that a certain percentage of polysaccharides against breast cancer, lung, liver, prostate and brain tumors is effective [2, 3]. The benefits of this product is promising to use this product in the diet. The growth period of this product consists of several stages and each of these stages requires different controlling condition [1]. The use of intelligent systems for automation in agriculture industries has been due to the development of computer systems and electronics in recent decades. With these systems, we can control the environmental parameters involved in mushroom production halls. Temperature is one of the parameters that shows a high impact on mushroom growth, and chemical reactions are intensive at higher temperatures. In biological processes such as growth, the effect of temperature can be easily observed where vast quantities of chemical reactions occur. The optimum temperature for mushrooms, depending on the stage and type of race, is 17 to 30 °C. The metabolism of consumed food by microorganisms in the compost contributes to their growth and activity, and as a result, it produces the heat. For example, rising compost temperature decreases crop production. Figure 1 shows the lack of mushroom production in the middle of the bed when the compost temperature is higher than the standard value [1].

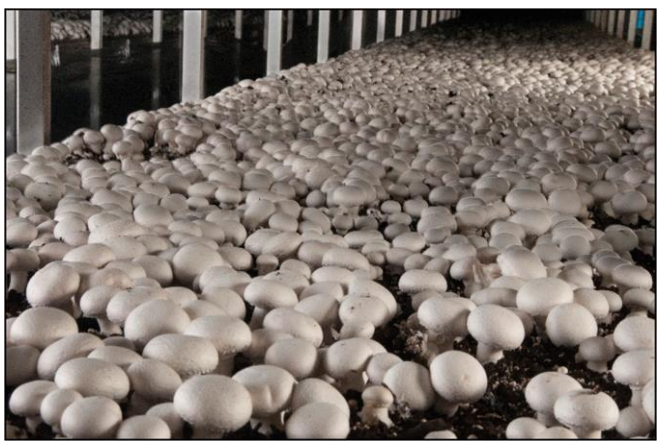

Fig. 1. The effect of temperature value on mushroom production

The successful cultivation of mushroom is possible when parameters such as temperature, humidity, and carbon dioxide concentration, pests and diseases and also preparing compost have been controlled and inspected properly. Environmental factors have the most influence on the quality of the product on the growth stage [4]. Problems in the field of parameters controlling on mushroom cultivation halls forced us to do studies on controlling these parameters. Manually and traditional controlling methods are under the influence of factors such as human, measurement and environment errors [1]. To resolve this problem, several studies were carried out with different control methods. Ardabili et al. [4] presented controlling system using fuzzy and digital controllers to control the environmental parameters of the mushroom production hall. Previous studies with presented methods have the complexity of calculation in control strategies. Today, predictive control is used in industrial applications to develop control strategies [5]. Among the systems that have the capability to model and predict the behavior of systems, can point to artificial neural networks. A neural network consists of a number of processing elements or computing nodes that are very simple and interconnected. This network is an algorithm information processing that is processing by dynamic response related to processing elements and their connections to lateral inputs [6]. The most common neural networks are Multi-Layered Perceptron (MLP) and Radial Basis Function (RBF) networks.

The main aim of this study is to present a predictive model of the temperature variation of the mushroom growing room by artificial neural networks based on the variables that are affecting on room temperature (ambient temperature, water temperature, fresh air dampers, circulation air dampers, and water tap). To reach this purpose, the study consists of three phases. The first stage is analyzing the required data. The second stage presents MLP and RBF models and the last stage presents the results and a comparison of networks and introduces the best model.

## 2  Materials and Methods

### 2.1  Data Collecting

This research was studied in one of the mushroom production halls of Sabalan agroindustry company (Sabalan Mushroom) in Ardabil province of Iran. The target hall has dimensions of 22, 5/6, 5/4 m (length, width, height, respectively) and has 1850 compost with a weight of approximately 15 kg for each compost. In order to data collecting operation, three PT-100 sensors were used. The location of temperature sensors in terms of height and width was in the middle of the hall height and width and in term of length were located on three points including the beginning of hall, the middle of the hall and the end of the hall. This method of the arrangement of sensors was used because the longitudinal air circulators in two top corners of the hall make airflow in transverse and height directions of Hall and provide thermal equilibrium on transverse and height directions of the hall, and if there is a temperature difference, this difference will be in the longitudinal direction of hall.

Compost generates heat. There is a need to maintain and stabilize the temperature of compost at every stage of the growth cycle and since the volume of compost is lower than the indoor air volume of the growing hall, the operation of stabilizing of compost temperature should be done by changing hall temperature.

Data collecting operations were done in the winter season and due to variations of ambient temperature, an external temperature logger was used to record temperature changes. Data collecting operations were performed at different reps. In order to record the required data, Autonics temperature controller TKM-B4RC was used that was equipped with RS485 output and related DAQ-master software. Ambient temperature, variations of circulation air dampers, variations of fresh air dampers, hot water tap and hot water temperature were as independent variables and indoor temperature of hall was as the only dependent variable. To adjust the room temperature and variation of input variables, it was using air conditioning systems (Fig. 2).

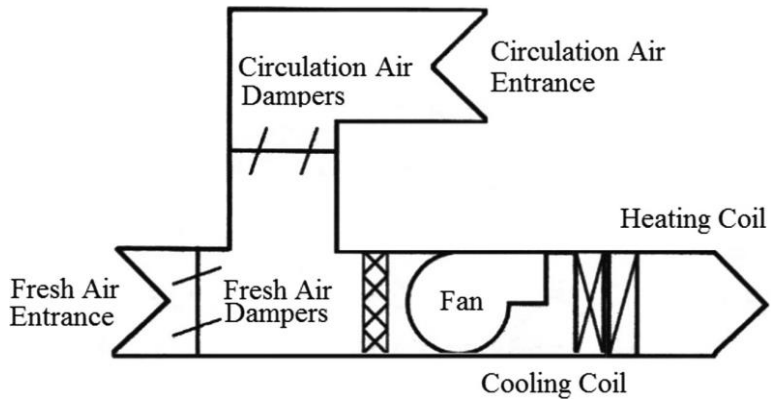

Fig. 2. Air conditioning system

This system is capable of cooling and heating, generating indoor air circulation by air ducts, providing required relative humidity and reducing the carbon dioxide concentration using fresh air dampers. Operation of temperature control in this system is performing by variating hot and cold water Debbie as well as opening and closing the air dampers. This system has two air inputs (Circulation and fresh air dampers) and an air output that the circulation air damper is for circulating the hall atmosphere and the fresh air damper is controlling temperature, humidity or carbon dioxide concentration by entering fresh air to hall [1]. Measurement of independent variables such as the air dampers and hot water tap were performed at three levels including minimum, medium and maximum value of actuators openness. According to the coolness of the air in the operation season, the outdoor air was used for cooling operations instead of cold water. Because the cold water would freeze and damage the coils in this season. Data were collected in 3 treatments and different repetitions to achieve high accuracy. Table 1 shows the different treatments for independent variables.

2.2 Artificial Neural Networks (ANNs)

Literature includes a vast number of machine learning methods used for the purpose of the modeling and prediction [7–26]. Machine learning models generally out-perform most of the statistical and mathematical models in term of computation cost, efficiency and accuracy [27–40]. ANNs are considered as efficient methods for developing reliable models. This study employs two types of neural networks including multilinear perceptron (MLP) and radial basis function (RBF).

Table 1. Treatments of independent variables for data collecting

| Parameter | Treatment | | |
|---|---|---|---|
| | Maximum | Medium | Minimum |
| Ambient temperature (°C) | −10 | 0 | +10 |
| Water temperature (°C) | 30 | 40 | 50 |
| Fresh air damper (Openness) | 1/3 | 2/3 | 3/3 |
| Circulation air damper (Openness) | 1/3 | 2/3 | 3/3 |
| Water tap (Openness) | 1/3 | 2/3 | 3/3 |

Before starting the training process, data have been divided into two categories of training data (with a share of 70%) and testing data (with a share of 30%), randomly. The training process was started with the different number of neurons in the hidden layer and the function of each parameter was measured with respect to the base parameter. To determine the optimal number of neurons in the hidden layer and to obtain the best predictor network, in first stage the network was trained with one neuron on a hidden layer. Figure 3 presents the structure of the RBF network.

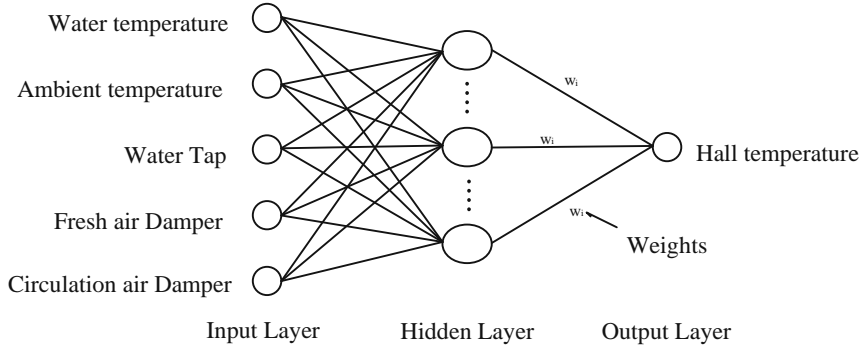

Fig. 3. The structure of RBF network

Evaluating the results have been conducted by employing Root Mean Square Error (RMSE), correlation coefficient (R) and mean absolute error (MAE) were used [41] to analyze the output of networks and target values.

$$MSE = \frac{1}{N}\sum_{i=1}^{N}(A-P)^2 \qquad (1)$$

$$RMSE = \sqrt{\frac{1}{N}\sum_{i=1}^{N}(A-P)^2} \qquad (2)$$

$$R = \left(1 - \left(\frac{\sum_{i=1}^{n}(A-P)^2}{\sum_{i=1}^{n}A_i^2}\right)\right)^{1/2} \qquad (3)$$

$$MAE = \frac{\sum_{i=1}^{N}|A-P|}{N} \qquad (4)$$

That A is the target value, P is the predicted values and N is the numbers of data.

## 3  Results

In this study, the temperature variation of the mushroom growing hall, as an critical factor of mushroom production, was modeled based on dependent variables including ambient temperature, fresh air damper, circulation air damper, water tap and water temperature using MLP and RBF networks. To perform modeling operations, there is a need to be aware of the general nature of the system that this is carried out by experimental data related to the system. For this purpose, the experimental data were obtained from the studied system using the introduced strategy in Table 1. Figure 4 indicates the results of data collecting and the relationship among the actuators and the related parameter.

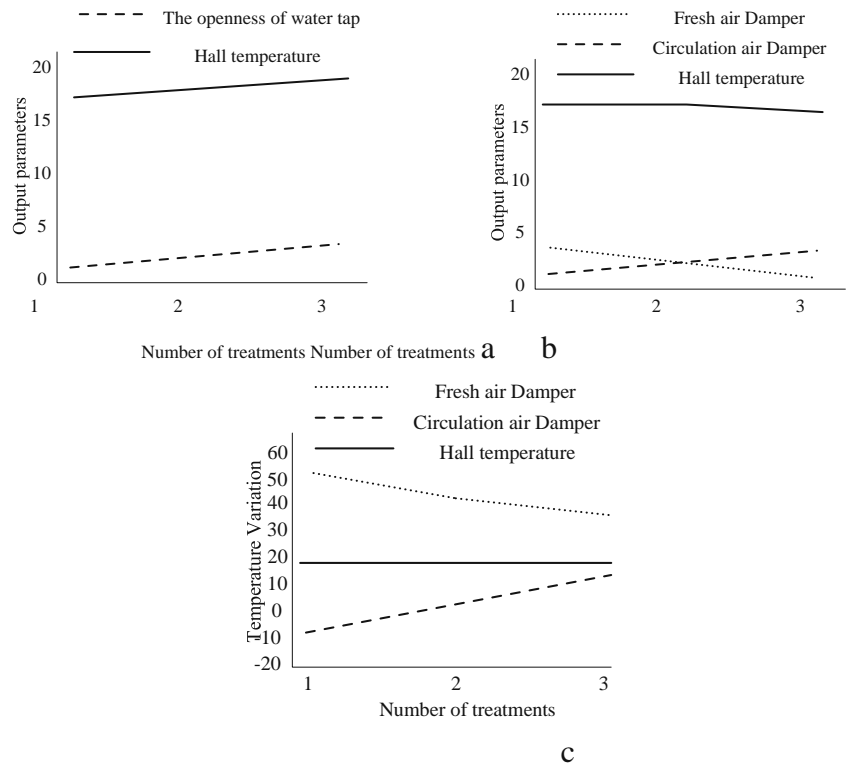

Fig. 4. The results of experimental data and the relation of dependent and independent variables

According to Fig. 4(a) by considering the fixed value of other parameters, opening hot water tap, increases the growing hall temperature. Figure 4(b) shows the variation of growing hall temperature by opening and closing air dampers when other parameters are fixed. Accordingly, if the rate of opening and closing of circulation and fresh air dampers be equal, respectively, the hall temperature almost will be fixed. In Fig. 4(c) by reducing the temperature of the water and by increasing the ambient temperature during the day, the growing hall temperature has undergone a constant trend.

### 3.1 Training Process

This section presents the results of choosing the best network for the training process based on the performance functions for RBF and MLP separately. Table 2 is related to MLP network, and Table 3 is related to RBF network.

Table 2. The result of selecting the best network for MLP model

| Numbers of the neuron on hidden layer/repetition | Value of performance function for validation data | Value of performance function for training data | Value of performance function for testing data |
| --- | --- | --- | --- |
| 12/1 | 0.55064 | 0.42202 | 0.84431 |
| 12/2 | 0.53321 | 0.41001 | 0.82541 |
| 12/3 | 0.52248 | 0.41056 | 0.84522 |
| Minimum value | 0.52248 | 0.41001 | 0.82541 |
| Maximum value | 0.55064 | 0.42202 | 0.84522 |
| Average | 0.53544 | 0.41419 | 0.83831 |

According to Table 2, the best result (lowest values of the performance function for testing data) is obtained in the second repetition of the training process. So the second network was selected as the best prediction network.

Table 3 was prepared to choose the best number of neurons in the hidden layer for RBF network.

Table 3. The result of selecting the best number of neurons on hidden layer for RBF network

| Number of neurons on hidden layer | RMSE | R | MAE |
| --- | --- | --- | --- |
| 4 | 0.2897 | 0.69 | 0.1016 |
| 8 | 0.1925 | 0.79 | 0.0675 |
| 12 | 0.1589 | 0.85 | 0.05575 |
| 16 | 0.1205 | 0.91 | 0.0422 |
| 20 | 0.0787 | 0.996 | 0.02746 |
| 24 | 0.0787 | 0.996 | 0.02745 |

As shown in Table 3, by increasing the number of neurons in the hidden layer, the correlation coefficient is increased and mean absolute error and root mean square error values are reduced.

After 20 neurons, these parameters remained stable and have not changed, so the number of 20 neurons in the hidden layer were selected as the optimal number of neurons. The networks were trained after selecting the optimal number of neurons to neural network Multilayer Perceptron and Radial Basis Function networks were trained. After the training process, test data were imported to developed networks and output data were generated to compare with the target values. The results of the target and the predicted values are shown in Figs. 5 and 6.

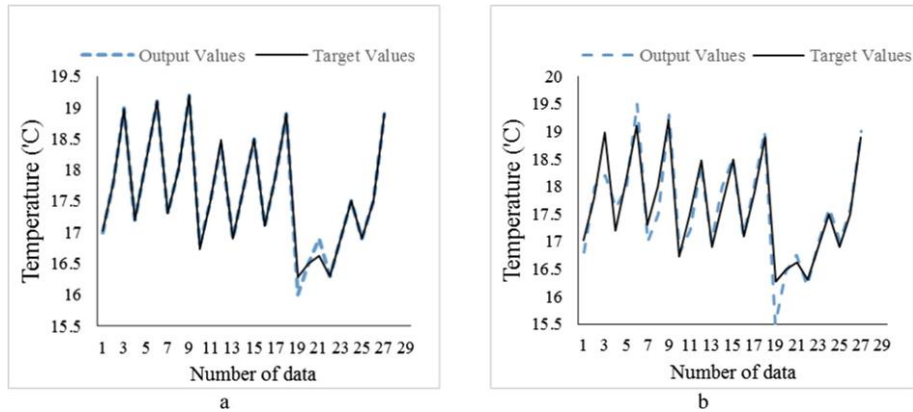

Fig. 5. Results of predicted and target values compared to target values. a) RBF network b) MLP network

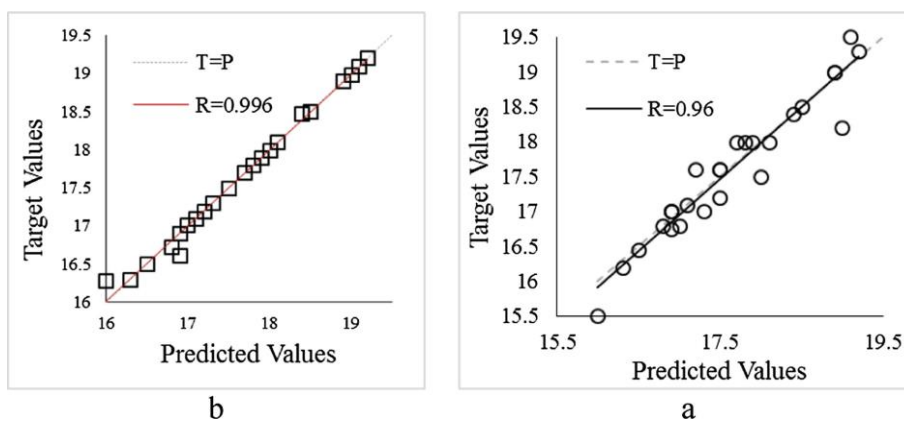

Fig. 6. Scatter plot of predicted and target values a) RBF network b) MLP network

According to Fig. 5.a, the predicted values of RBF network are following target values well and have less deviation and error from the target value, but predicted values of MLP network (Fig. 5.b) have a large error and deviation from target values and has lower compliance with target values compared to RBF network.

Based on the results of Fig. 6 and according to the description mentioned about correlation coefficient, it can be said that the output values of RBF network have 99.6% of linearship and the output values of MLP network have 96% linearship with target values. To display these results as a statistical factor, the output of models were compared with target values using the comparison parameters that were mentioned in Materials and methods. The obtained results were tabulated in Table 4.

Table 4. The results of comparison parameter for two types of networks

| Network type | MAE | RMSE | R |
|---|---|---|---|
| MLP | 0.137 | 0.9085 | 0.9612 |
| RBF | 0.02746 | 0.787 | 0.996 |

According Table 4, the results of comparison parameters indicate that the results of RBF network have high correlation coefficient (0.996) and low RMSE and MAE values (0.787 and 0.02746, respectively) compared to MLP network. Due to the high correlation coefficient for RBF network (0.996), It can be said that process modeling, compliance and linear correlation predicted by the RBF network is higher than MLP network that is confirming the obtained results from Figs. 5 and 6.

On the other hand, RBF network with the lowest root mean square error (0.787) and the lowest mean absolute error (0.02746) generated closest predicted values with minimal errors compared to the MLP neural network and it can be said that RBF has high ability to model the temperature variations compared to MLP network in this study. Therefore the designed model based on RBF is predicting the temperature value more accurate with low deviation to target values compared to the MLP network.

Ardabili et al. [42] developed a fuzzy modeling system in order to predict the temperature of the mushroom growing hall that the correlation coefficient and mean absolute error between the predicted and target values were calculated 0.67 and 0.232, respectively. The present study indicates the improvements in the prediction of temperature variations using artificial neural networks. One of the reasons that led to this happening, is that the fuzzy systems unlike the artificial neural networks, is operating by the defined laws. These rules can be affected by the accuracy of laws defining and can have a negative effect on system precision.

Figure 7 presents the error values for the predicted values of each network from the desired values. The zero value of deviation is related to the Target value. The blue line indicates the deviations related to RBF network and the red line indicates the deviations related to MLP network. Based on Fig. 7, MLP network has the maximum deviation from the target value.

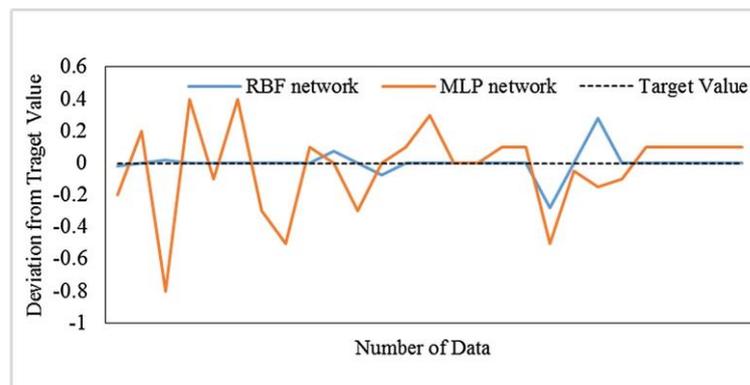

Fig. 7. Deviation Of predicted values of networks from the target value

According to Fig. 7, if the output of networks is compared in the same input values, it can be said that the deviation of temperature from target values in RBF network is higher than MLP network. This means that the energy losses of MLP network are higher than RBF network. This energy losses on MLP network can be reduced by the changes that can be applied in network inputs. This losses of energy is equal to increasing the failure risk of the system on MLP model compared to RBF model.

## 4  Conclusion

This study is performed in a mushroom growing hall with the aim of modeling of temperature variations. Accordingly, modeling systems including MLP and RBF networks was used. The results of the data collecting process reflected the dependence of temperature value to independent variables. Therefore, results were prepared after the modelling process and extracting the output values of networks and comparing them with target values. This results showed that the RBF network has high accuracy and better performance compared to MLP network and also using RBS network will reduce energy consumption, system failure, and costs. Thus, the neural network with radial basis function was chosen as a predictive network of hall temperature in this study. For the future works, more sophisticated machine learning methods must come to consideration, e.g., [42–51].

Acknowledgments. This publication has been supported by the Project: "Support of research and development activities of the J. Selye University in the field of Digital Slovakia and creative industry" of the Research & Innovation Operational Programme (ITMS code: NFP313010T504) co-funded by the European Regional Development Fund.